# THE IMPACT OF NEW TECHNOLOGIES IN PUBLIC FINANCIAL MANAGEMENT AND PERFORMANCE:
*AGENDA FOR PUBLIC FINANCIAL MANAGEMENT REFORMANCE IN THE CONTEXT OF GLOBAL BEST PRACTICES*

Prof. Amos DAVID
Nancy University, Nancy FRANCE
Research team: SITE-LORIA[1] (Modeling and Development of Economic Intelligent Systems)

---

[1] http://site.loria.fr ; http://www.loria.fr/~adavid

# TABLE OF CONTENT

## I. INTRODUCTION

Information and Communication Technologies (ICT) has practically penetrated into all spheres of life. Therefore a closer look at the impact of ICT in public financial management and performance is highly justified.

For a better understanding of the discussions in this paper, it is necessary to first clarify the semantics of some of the main concepts of the topic. The presentation in this paper is voluntarily non technical.

In this paper, the term "new technologies" is limited to ICT. Furthermore, we restrict our presentation to only the functionalities of information systems without delving deep into the infrastructures for the ICT. The choice is justified because the topic is expected to focus on the impact of the new technologies and not on the understanding of the technologies. Also for this reason, we first present what we understand by *public finance, management* and *performance*.

**Public finance**[2] is defined as a field of economics concerned with **paying** for *collective* or *governmental* **activities**, and with the **administration** and **design** of those activities. The field is often divided into questions of what the government or collective organizations should do or are doing, and questions of how to pay for those activities.

**Management**[3] in business and human organization activity, in simple terms means the act of getting people together to accomplish desired goals. Management comprises planning, organizing, resourcing, leading or directing, and controlling an organization (a group of one or more people or entities) or effort for the purpose of accomplishing a goal. Resourcing encompasses the deployment and manipulation of human resources, financial resources, technological resources, and natural resources.

In the field of management, **performance** is the ultimate result of all the efforts of a company or an organization. These efforts are to do good things, good way quickly, at the right time, at the lowest cost to produce good results that **meet the needs and expectations** of customers, their satisfaction and achieve the **goals set by the organization**.

Since the measure of performance within the framework of this paper concerns the activities proposed by collective or governmental organizations, we present below how activities can be viewed so that the impact of ICT can be well assessed. Activities will be viewed as services or more precisely as public services.

We believe that there is need to consider performance from the perspective of *effective performance* and the *perceived performance*. In fact the real or effective performance might not correspond to the perceived performance. We also need to specify the identified success factors for the use of ICT in general, the common advantages and the common limits for the use of ICT.

## II. SUCCESS FACTORS, ADVANTAGES AND LIMITS OF THE USE OF ICT

Two major success factors for ICT in public services are the *ICT infrastructure* and the *end-users ICT literacy*. The use of ICT and their impact can be felt only if the ICT infrastructure is available and the end-users are ICT literate. Here we refer to ICT literacy as the basic understanding of the use of the common functionalities of the tools available on Internet for information exchange. In this paper, the ICT infrastructure concerns mainly the Internet network and the accessibility facilities for the end-users of the Internet.

---

[2] http://en.wikipedia.org/wiki/Public_finance, September 2008
[3] http://en.wikipedia.org/wiki/Management, September 2008

***Common advantages of the use of ICT***

We present below four common advantages of the use of ICT in all sectors and illustrate them with an existing public service - *the passport and visa services[4]*. This public service exists since the inception of control across country borders. The process of passport or visa has ever been carried out manually. So, introducing the ICT in the process illustrates perfectly their impact on public service and thereby allows the measure of the performance of the service with the use of ICT compared to the traditional process without ICT.

*Time saving: (transport, delivery time, response time, etc.)*

The main time saving factor is the flexibility of scheduling when to engage on a process. For example a visa application form can be filled when the applicant has a free time and not necessarily during the opening hours of the Consulate. The data filled are verified immediately for validity. Acknowledgment is sent immediately as prove of application. Payment is made online with credit card thereby saving the time of passing through the cashier.

*Money saving: (transport cost, material cost, environmental cost, etc.)*

The applicant doesn't need to travel to the Consulate two times (first time for collecting the application form and booking for appointment, second time for submitting the application form and for physical presentation). He needs to go to the Consulate only once since the application form would have been filled online. The cost for traveling the first time is here saved. Also the potential wastage of paper is removed. For example, it is a common practice to make a photocopy of the original application form for a test. When the form has been correctly filled, the original form is then filled. In the case of online application, no paper is printed and corrections are proposed online. Finally, reducing paper consumption helps improve the quality of the environment – reduce forest destruction since paper is produced from wood, and reduce waste processing.

*Improved security: (transportation, transactions, trace of activities, etc.)*

The reduction of the number of transports necessary for processing a visa contributes to reduction of dangers and insecurity linked with transportation. For example, the reduction of the number of transports will reduce traffic and thereby reduce the potential of road accident. Also in countries where road transport constitutes some form of danger by arm robbers, reduction of the number of transports will also reduce exposure to arm robbery. Online transaction for payment will contribute towards two types of security – the reduction of exposure to arm robbery and the reduction of temptation to bribery. It is in fact a common believe that arm robbery succeeds since the arm robbers are sure of getting money during their operation since the majority of the population has no other means but carry raw cash on them when travelling. Also, since those practicing bribery rely on the direct contact with the client (or applicant), coupled with the absence of trace of their transaction, the fact that the trace of all transactions is recorded helps dissuade the practice of bribery. Transactions for payment are made online and acknowledged with prove of payment.

---

[4] https://portal.immigration.gov.ng/index.htm

### *Common limits of the use of ICT*

Some of the limits of the use of ICT are linked with the financial implication of putting them in place and the inherent functionalities of the ICT.

*Financial investment: (public Internet network, private work stations)*

One of the main problems of the use of ICT in public service is the required financial investment necessary to build and operate them. Not only is it necessary to build them, they should also be made available to end-users. Unfortunately, most end-users cannot afford to purchase a personal computer with access to Internet. Even for some potential end-users that have personal computers, the cost of access to Internet services is still prohibitive for their budget. Most of the developed countries have national plans for global and generalized ICT infrastructure development. Recent example is that of Franc's 2012 "Global numerical plan". It is also of high interest to note that the Federal governments of Nigeria, as well as some states in Nigeria have projects of establishing generalized ICT development[5].

*Security on the control of transfer protocols: (hackers, spammers, etc.)*

The fear of interception of data transferred over the Internet is wild spread. This has led to some end-users to avoid using Internet. For example, the amount of payment transaction over Internet is highly dependent on this type of fear. However, there has been a lot of progress to make data transfer over Internet as secure as possible. The other security problem is as a result of the ravage done by hackers and spammers. The hackers work on bypassing all security devices on the Internet while the spammers make use of the democratic nature of access to Internet, which is the absence of total control.

### *Positive side effects of the use of ICT for the process of visa*

One of the major side effects of the use of ICT for visa processing is the fact that data filled are in electronic form. Therefore, they can be used to populate a database, which is a means of observation by the administrators of the public services as we will explain in the next section. The information stored in the database can be used to provide tangible information not only to the immigration service but also to the economic and tourist sectors. For example, statistics could be provided from business trips for the business sector and duration of tourist visits for tourist organization. Distribution over the year of visits could be obtained. And finally, the yearly trend of the different categories of visit can be obtained. These are vital indicators for planning and development.

Now that we have summarized the basic success factors, some advantages and limits of the use of ICT for enhancing public service, we present in the next section the fundamental principles underlying the success or the failure of the use of ICT for public services.

---

[5] http://odili.net/news/source/2008/oct/30/28.html: The Guardian, 30 October 2008: LAGOS State Governor, Babatunde Raji Fashola, has approved the creation of Information Communication Technology (ICT) units in all ministries in order to provide improved service delivery in all areas of governance in the state. According to a statement, the units, which will come into operation before the end of November, would, among other things, serve as help desks, as well as provide IT support to the ministries.

## III. THE EFFECTIVE AND THE PERCEIVED PERFORMANCE OF PUBLIC SERVICES

We have heard the following expressions more than once from the population:
- The government is not doing anything;
- We don't know how our money is spent;
- A lot has been spent but we can't see the result;
- The priority of the government is misplaced;

Most of these remarks can be traced to the inherent problems associated with public finance and management performance as defined in section I - the **design** and **administration** of services. The environment of a public service can be represented with the following schema.

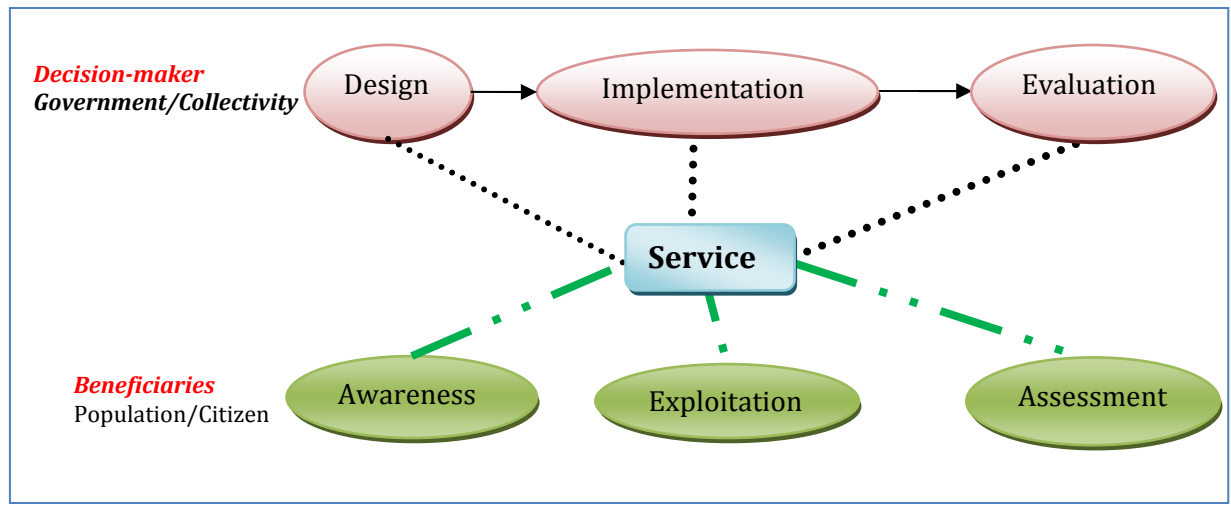

*Fig I.   Services in a social organization*

As illustrated in fig I, a service can be considered from the perspective of the decision-maker, who in our case, could be a government or a collectivity or the human actors within these organizations. ICT can be employed in the three phases that concern the decision-maker - *design, implementation and evaluation*.

The beneficiaries of a service can employ ICT in any of the three phases - *awareness, exploitation and assessment* - for guarantying a high level of efficiency. We present in the following sections each phase in the environment of a service and illustrate how ICT can be employed in order to improve the end-result of each one of them. We believe that a high efficiency of each phase will produce a high global efficiency.

We will illustrate the discussion in this paper using the example (or case study) of a government organization, such as federal, state or local government, in which the administrators are concerned with the agricultural environment of the region administered. Therefore, in this environment, the main human actors are the ***administrators***, the ***farmers*** and the ***other members of the population*** of the region. The presented principles apply to any other socio-economic sector of a nation. In section II, we gave an example of the application of ICT in an existing service. In this last example, we can imagine that the service has not been defined and consequently ICT have not been applied.

## The phase of design of a service by the decision-maker

The design phase of a service is of high importance in the sense that a service cannot exist without this phase. The service has to be **imagined** with a **set of goals** for a defined **population**. It is with regards to these activities that ICT can play important role.

An important question to ask is "how is the idea to put up a service burn"? It should be noted that the idea of putting up a service cannot arise without identifying a population's need or a subset of this population. So we can say that the need for a service arises from the consciousness of a population's need and the association of this need with the appropriate service. The population's needs may be considered as problems for that set of population. It is the responsibility of the decision-maker to transform these problems into stakes – that is problems for which solutions should be provided. This underlines the importance of human factor in the effectiveness of a service.

From our case study, let us consider six of the activities of a farmer
- Prepare the land (for example plowing and harrowing),
- Plant some crops,
- Maintain the plantation (for example through weeding, application of fertilizer, application of herbicide),
- Harvest,
- Conservation,
- Market the farm produce.

If we consider the role of the administrator in charge of agriculture in the government structure in our case study, we expect that one of his main tasks would be the identification of the farmers' problems as related to the farmers' activities. The administrator may adopt one of the following attitudes:
a) Refuse to identify the farmers' problems through observation, that is refuse to observe the farmers' problems.
b) Observe the farmers' problems; perceive the problems but for one reason or another doesn't consider them as problems for which solutions should be sought.
c) Perceive the farmers' problems and consider that the problems should be solved.

In situation (a), unfortunately there is nothing that ICT can do to assist and consequently ICT will have no impact on public service since there will be no service proposed to start with.

In situation (b), ICT can be applied for the process of observation. For example, the traditional process of observation is through interviews and questionnaires, providing a forum for the farmers to express their perceived needs. The expressed needs can be complemented with the expertise of the administrator through experiences gained in the past and/or from other regions. ICT can be employed for the three traditional procedures for observation. The interviews could be conducted through electronic mails or through discussion groups. Questionnaires can be posted and filled through forms on Internet. And finally, the past experience of the administrator could be documented and made available through web pages or any other form for making information available on Internet.

A more advanced use of ICT could be to perform cross analysis on the collected information. This type of analysis reveals some hidden phenomenon that could constitute the root of some problems. This point will be developed in section IV.

Unfortunately, even if the administrator becomes aware of the problems, he might, for one reason or another, fail to transform them into stakes, that is, problems to solve. One of the main reasons for this attitude is the fact that the administrator may not stand to lose anything in the absence of any decision for solving the problems. In other words, he might ignore or neglect what he stands to gain or lose leaving the problems unsolved. Here again, the impact of ICT in the building of the service is dependent on the human factor.

Using our case study, assuming that the administrator observed the following problems encountered by the farmers for the various activities and decides to set up a public service to solve them. This corresponds to situation (c) above.

The identified problems could be the following:
1. Prepare the land (for example plowing and harrowing): The administrator noticed that the farmers are not aware of the opportunities provided by the government to facilitate land preparation. These opportunities might include provision for tractor rent and the related information such as the period and financial implications. One of the complaints of the farmers might concern the inability to plan when to book for a tractor which in turn reveals the problem of weather forecast since land preparation is determined by the climate.
2. Planting crops: The administrator might have noticed the farmers' problems in obtaining seedlings and the adaptation of the crops to their soil type. Related problem could also be on lack of information on the evolution of market demand for some specific crops.
3. Maintain the plantation (for example through weeding, application of fertilizer, application of herbicide): Problems here could concern the adaptation of herbicide to the crops planted, where to obtain the herbicide, and cost evaluation of the use of a specific herbicide. Other problems could concern the adapted fertilizers to use for some specific crops. It could also concern a comparative study of "biological" style of maintenance which concerns maintenance of plantation with the use of neither herbicide nor fertilizers.
4. Harvest: Some problems might concern the harvest techniques and the available politics to facilitate harvesting of crops. For example, harvesters might have been made available by the government but completely ignored by the farmers.
5. Conservation: Most farmers are highly concerned with conservation techniques. They don't know the existing techniques, the costs and how to implement them.
6. Market the farm produce: In some cases, the government might have adopted politics in favor of commercialization of farm produce which the farmers might have ignored.

It should be noted that the information provided does not constitute lectures in agriculture but providing information for specific and identified agricultural problems.

### The phase of implementation

The phase of implementation follows the phase of design in which the services to put up would have been identified. ICT is very suitable for this phase. Most of the enumerated problems above can be solved by providing adequate information. For example, a portal can be created to provide the following information:
1. Information related to tractor rent, Booking service for tractors, Meteorological information
2. Information on seedlings, soil types for the crops, market state for the crops
3. Information on weeding techniques, adapted fertilizers for the crops, the adapted herbicide for the crops, price and consequences, and techniques for plantation without fertilizers and herbicide
4. Information on harvest equipments on rent, cost, techniques, booking, etc.
5. Information on large or micro conservation techniques, costs, available politics, etc.

6. Information on the politics to encourage and facilitate marketing of farm produce.

### The phase of evaluation

The evaluation phase is also very adapted for ICT use. In the evaluation phase, the decision-maker who in this case would be the administrator, needs to have feedback from the farmers, so as to assess the adequacy of the proposed services to the farmers' needs. The feedback can be obtained through questionnaires, group of discussion and through frequently asked questions.

There is also a possibility of implicit automatic measure of the adequacy of a service through the measure of visits to a particular service. A simple measure is to set up a counter to measure the frequency of visit. A more sophisticated measure allows the measure of who make use of what and not just counting the number of access to a particular web page.

While the three precedent phases concern the decision-maker, the phases below concern the beneficiaries of the proposed services.

### The phase of awareness

It is evident that a service cannot be employed if it is ignored, that is, the potential user is unaware of the existence of the service. ICT can be used to make known the proposed services. For example, one of the elementary functionalities of Internet is the use of search engines. The portals for the services could be indexed by the search engine and thereby make them known to the potential beneficiaries.

As indicated in section II, one of the main problems is the beneficiary's ICT literacy. For example, it would be very difficult to rely on the assistance of others each time the beneficiary is in need of one of the functionalities of ICT.

Another means of making the services known to the potential beneficiaries is through "traditional" means, for example through brochures.

One other important point is that the potential beneficiary must be motivated to find solution to his problem. Just as it applies to the decision-maker who must identify a problem and transform it into stake, the beneficiary must transform his problem into stake, thereby transforming the problem to problem that must be solved.

### The phase of exploitation

It is in the exploitation phase that there is need for ICT literacy. This will dictate the degree of independency of the beneficiary.

There are still some psychological barriers hindering the various functionalities of ICT for public services. As mentioned in section II, some potential beneficiaries are still reluctant at using the ICT because of the fear they exercise on security. For example some people refuse to make payment using ICT.

A mot technical problem concerns the use of the information obtained. In fact some information might have been well preprocessed and structured, making it directly usable by the beneficiary. In some other cases, the beneficiary might have to perform restructuring before obtaining the form exploitable for their need. This is often the case when a list of information is provided. For example, yearly data may have no meaning unless transformed inform graphics. But before obtaining the graphics, the raw information obtained has to be transformed into numerical information for the graphic presentation. This operation cannot be performed by some beneficiaries.

### The phase of assessment

The assessment phase concerns the beneficiary's point of view on the service provided. This is generally measured compared to his information need.

There is need to pay special attention to this phase since it is directly connected to the phase of design by the decision-maker. This explains why the potential beneficiary should be involved when the service is designed either through the observation of their need or through the past experience of the decision-maker.

It is in this phase that there could be a discrepancy between the real performance and the perceived performance. The perceived performance may not correspond to the real performance since the functionalities of the service might not correspond exactly to what the beneficiary expects.

A formal way of making the real and the perceived performance to correspond to each other is to provide the potential beneficiary with the mean of specifying the functionalities he expects from the service. This is however seldom practiced because of the heterogeneous nature of specification. The adopted methodology consists in proposing the services based on observation of the activities of the potential beneficiaries and transform their problems into stakes for which ICT can contribute towards their resolution. This means that the quality of observation determines the distance between the perceived and the real performance of a public service.

## IV. THE ECONOMIC INTELLIGENCE APPROACH FOR THE USE OF ICT IN PUBLIC SERVICES

Below are two definitions of the term "Economic Intelligence" (EI):
[1] It is all the coordinated actions of collection, processing and distribution of useful information for the economic actors with the aim of its exploitation. These actions are taken legally with all the guarantees of protection necessary for the conservation of the company's patrimony, in the best conditions of quality, of delay and of cost [10].
[2] It is the process of collection, processing and distribution of the information that has as goal the reduction of uncertainty in strategic decision making [11].

Indeed the more and more dominant role of information in the socio-economic world and generally in the organizations is not to demonstrate any more. Information and communication technologies, more particularly the data processing tools and the Internet, allow managing information of different natures: primary information, secondary information, tertiary information, information with added value. While the primary information is the direct work of the producers, the secondary information and the tertiary information are transformations of the primary information, to feed databases (or information bases). The value-added information is the product of analysis and synthesis of these various types of information. Information is more and more used as reference object and as aid tool for making decisions of strategic order. The concept of EI asserts itself where it is a question of studying the processes involved in the production of the interpretable indicators for decision making based on internal and external information in an organization.

### EI and watch process

IE process is based on the **process of watch**. We distinguish here two types of watch: **tactical watch** and **strategic watch**. The tactical watch feeds the field operators of the company with information, and the studied timescale concerns the present and the very short term. The strategic watch is characterized by the distribution of information to the management entities of the company. The studied timescale is the present, the very short term, the middle term and **long term**. The following stages represent the process of watch.

(a) Identification of needs in the form of problems to solve or stakes *(threat, risk, danger)*,
(b) Identification of the types of result,
(c) Identification of the types of necessary information to obtain the result,
(d) Identification of the relevant information sources,
(e) Validation of the information sources,
(f) Collection of information,
(g) Validation of the information collected,
(h) Processing of the information collected for the calculation of indicators,
(i) Interpretation of the indicators,
(j) Decision making for the resolution of the identified problem**.**

The process of watch appears as a succession of the above stages with a possibility of their iteration. In EI context, the stages (a), (i) and (j) are of particular importance. Indeed EI process is a global process where the orientation chosen in every stage will determine the type of the final result. For example, if a company decides to make decisions with offensive intention related to a competitor, the necessary information, the sources of this information, its processing and the interpretation of the final result will determine the scope of the company's decision. It should be noted that the necessary information is not only factual information as in databases of company but is of diverse nature as by its source, its validity, their scope, and informal because it may not have been intentionally published.

### Architecture of an extended information system

The following figure shows the importance of an information system (IS) in EI process. We present the figure by the arcs that connect the elements of the diagram:

- **Selection**: It allows the development of the company's IS that can be i) the database of production (commonly used for the company's organization), ii) all the information support for an Information Retrieval System (in documentation for example) or iii) a strategic information system based on data warehouse. This IS is built from heterogeneous sources of information by means of a filter of the reality.
- **Matching:** Matching allows any type of user to access the information of the IS. Two main methods of information access are presently proposed to the users: access by **investigation** and access by **query**. The investigation method is based on the technique of hypertext. The queries are expressed by means of Boolean operators. The result of matching is a set of information.
- **Analysis**: Techniques of information analysis are applied to the result in order to add values to the collected information. For example, the assistant of a department head whom we consider as a watcher can establish a benchmark for his department head. So, reports supplied by the assistant who knows well the aims of the boss will be a good base for making a decision. Presently available open IRS does not provide this functionality.
- **Interpretation**: This concerns the decision-maker to make good decisions. One sees here all the interest to acquire knowledge on the decision-maker, integrated into a data warehouse and used to build data marts that are specific to a group of decision-makers or better still to a particular decision-maker.

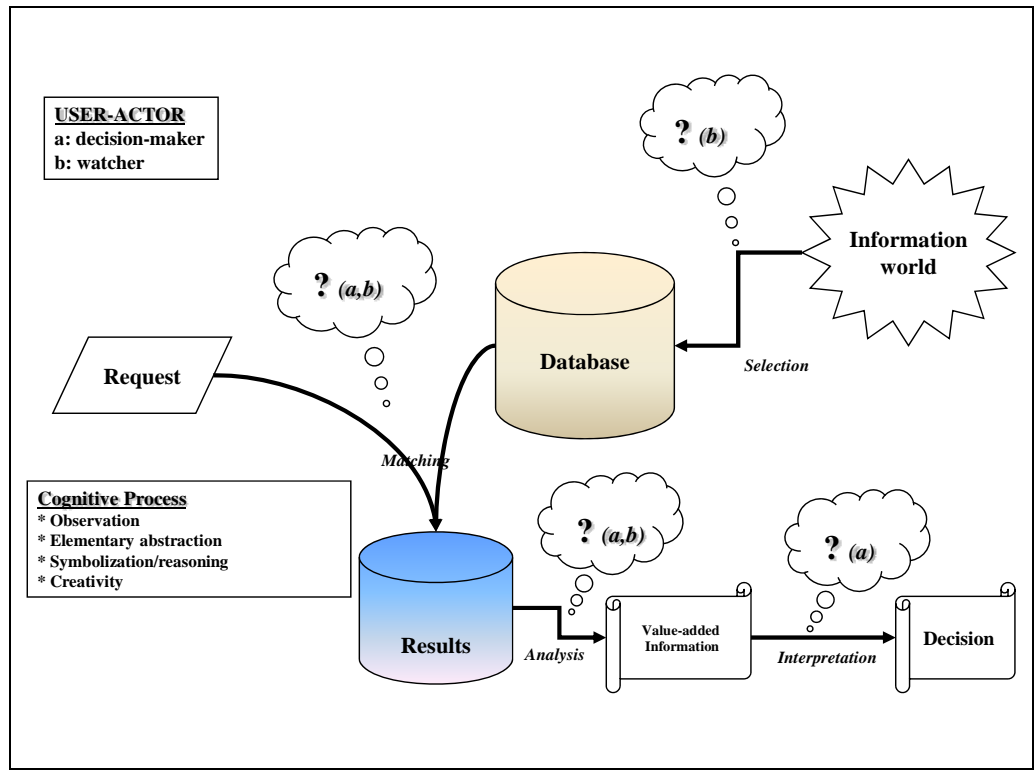

*Figure 1: Architecture of an Extended Information System*

## Comparison of Watch and Economic Intelligence

We present in the table below the differences between the task of watch and that of EI as presented in [w02].

| Watch | Economic Intelligence |
|---|---|
| From the information will come the strategy | From the strategy will come the information need |
| Collect the maximum of possible information | Search for minimum necessary information |
| Acquisition of knowledge | Assistance for decision making |
| Task for specialists | Task for generalists |
| Exhaustive information retrieval on a subject | Consultation of a sample of information |
| Accumulation of information | Elaboration of information |
| Data (a lot of "white" and repetitive information) | Investigation ("grey" and diverse information) |
| Information of rather static character | Information of rather dynamic character |
| Observation and anticipation techniques | Legal offensive and positioning strategy |
| Direct approach and concentrated efforts | Indirect approach and dispersed efforts |
| A stake of the company | A stake of a network of actors or of national interest |

Economic intelligence approach can be used in decision making process in some of the following aspects for solving a decision problem.

### Stake identification through indicators

- Identify new opportunities and problems; monitoring the socio-economic environment
    - Economic indicators
    - Social indicators
    - Educational indicators
    - Transport indicators
    - Cross analysis of indicators

### Constitution of Directory of strategic information sources

- Federal government ministries
    - State governments ministries
    - NISER
    - Central bank
    - Chambers of commerce[6]

### Use ICT to feed governing bodies with information

- Frequently asked questions
- Comments
- Critics
- Difficulties
- Suggestions
- etc.

## V. EXISTING INITIATIVES

Some initiatives have been taken either by the federal government or by the state government all around the world. The initiatives aim at providing information to socio-economic actors as well as to the population at large.

In this line, information is made available on databases and in some specific cases; the database contains strategic information, such as strategic plans, trends.

Below are some specific initiatives in France and in Nigeria.

### Initiatives in France
- National Institute for Statistics and Economic Studies[7]
- Ministry of Finance[8]

### Initiatives in Nigeria
- Federal Ministry of Finance[9]
    - Economic performance reports
- Nigerian Institute of Social and Economic Research (NISER)[10]

---

[6] http://www.lagoschamberng.com

[7] http://www.insee.fr/en

[8] http://www.budget.gouv.fr ; http://www.comptes-publics.gouv.fr

[9] http://www.fmf.gov.ng

[10] http://www.niser.org/hfo.htm

- State governments web sites
  - Kogi state[11]
  - Lagos state[12]
    - *Comprehensive Infrastructure Master Plan for Draft Master Plan Report Lekki Sub-Region – Lagos – Nigeria*

## VI. CONCLUSION

As we have seen with the two illustrations in section II and III, ICT can be harnessed whenever it is possible to de-materialize an object or an activity. This explains why ICT has penetrated all spheres of live.

The impact of ICT in public financial management and performance can be felt in all the six phases concerning the decision-maker and the beneficiaries of public services, from the design phase by the decision-makers to the assessment phase by the beneficiaries.

There are already existing initiatives of deploying ICT in existing public services. However new public services can be invented using the same approach as exposed in section III. Practically all socio-economic organization can deploy ICT for their activities such as secretariats, universities[13], hospitals, electronic commerce, electronic governance, banking, etc.

---

[11] http://www.kogistateofnigeria.org
[12] http://www.lagosstate.gov.ng/web/lagos/home
[13] See : ICT centers in the Nigerian universities, implementation of numerical working environment